\documentclass[smallabstract,smallcaptions]{dccpaper}

\usepackage{epsfig}
\usepackage{citesort}
\usepackage{amsmath}
\usepackage{amssymb}
\usepackage{color}
\usepackage{url}
\usepackage{booktabs}
\usepackage{caption}

\usepackage{fancyhdr}

\newlength{\figurewidth}
\newlength{\smallfigurewidth}

\begin{document}

\title
{\large
\textbf{A Low-Complexity Speech Codec\\
Using Parametric Dithering for ASR}
}
\author{%
Ellison Murray, Morriel Kasher, and Predrag Spasojevic\\[0.5em]
{\small
\begin{minipage}{\linewidth}
\centering
\begin{tabular}{c}
WINLAB, ECE Department, Rutgers University\\
\{ellison.murray, morriel.kasher\}@rutgers.edu, spasojev@winlab.rutgers.edu
\end{tabular}
\end{minipage}
}
}

\maketitle
\thispagestyle{fancy}

\vspace{1 mm}

\begin{abstract}
Dithering is a technique commonly used to improve the perceptual quality of lossy data compression. In this work, we analytically and experimentally justify the use of dithering for ASR input compression. We formalize an understanding of optimal ASR performance under lossy input compression and leverage this to propose a parametric dithering technique for a low-complexity speech compression pipeline. The method performs well at 1-bit resolution, showing a 25\% relative CER improvement, while also demonstrating improvements of 32.4\% and 33.5\% at 2- and 3-bit resolution, respectively, with our second dither choice yielding a reduced data rate. The proposed codec is adaptable to meet performance targets or stay within entropy constraints.
\end{abstract}

\Section{Introduction}
\vspace{-2mm}

Quantization is fundamental to analog-to-digital conversion (ADC) \cite{gray1998quantization} and underlies lossy audio compression methods such as waveform, subband, and transform coding \cite{vanderkooy1987dither,brandenburgMP3}. This process introduces distortion, or quantization error, which can be made less perceptible through dithering.

Dithering extends beyond perceptual coding, having applications in machine learning: federated learning \cite{lo2021architectural} and automatic speech recognition (ASR), and in graphic design (halftoning) \cite{Freitas2011}. For ASR, frequency-selective dithering has been studied as a means of compensating for the spectral feature distortion of MPEG-3 coding \cite{borsky_spectrally}.

Previous results have been limited to outdated ASR acoustic models based on Gaussian Mixture Models and early Deep Neural Networks \cite{borsky2017dithering}, without consideration of critical factors such as implementation complexity and entropy trade-offs. Moreover, there remains a limited rigorous justification for dithering within ASR systems, with primarily black-box evaluations \cite{borsky2015advanced} and minimal exploration of its effect on low-level uniform quantization, all of which we aim to address.

Implementation constraints are particularly relevant for enabling real-time offloading of speech data from low-powered wearable electronics \cite{ha2013towards}. Standard codecs such as Opus provide highly efficient, near-lossless compression even at low rates, but rely on FFT-based computations \cite{vos2013opus}. A low-complexity speech codec would improve the functionality of speech-interactive wearables, reducing communication barriers for users with hearing impairments or in multilingual environments.

We propose a low-complexity parametric dithering quantization system designed to balance ASR performance and implementation efficiency. We hypothesize a link between ASR accuracy and the total power and the autocorrelation properties of quantization error. We then introduce two adaptable dither distributions that allow us to explore this trade-off. Through experimental evaluation, we investigate the effectiveness of these distributions and present results that support our framework as a step toward a principled understanding of dithering in ASR.

\Section{Background}
% \vspace{2mm}

\subsection*{Dithered Quantizers}

We consider a $b$-bit uniform scalar quantizer with finite resolution ($b\in\{1, 2, 3\}$), operating under a low entropy constraint. While non-uniform quantizers achieve lower distortion at low rates \cite{gray1998quantization}, we prioritize the simplicity of uniform quantization for practical implementation. We define this mid-rise quantizer function $Q(.)$ with codebook value output, $C_k = k\Delta+\Delta/2$, and decision boundaries, $T_k = k\Delta $, where $k \in \{1, 2, \cdots, 2^b\}$, with $k$ denoting the index of each quantization output bin and $\Delta$ denoting size of the bin. 

A uniform quantizer, designated as such by $C_{k+1}-C_k = T_{k+1}-T_{k} = \Delta$, can also operate in a mid-tread configuration ($C_k = k\Delta , \quad T_k = k\Delta  -\Delta/2  $). We adopt the mid-rise configuration, which ensures that the output is centered around zero: a crucial property at low bit depths.

We consider an analog input and dither samples,  $x \in X$ and $v \in V$, where $X$ and $V$ are the respective random variables; the quantizer input sample $y \in Y$ is the sum of our input and dither samples $x+v = y$, with random variable $Y$. The quantizer output is given by $Q(y) \in \{C_1, C_2, \ldots, C_{2^b}\}$,  where $Q(y) = C_k \implies T_k \leq y \leq T_{k+1}$
The error of our dithered quantization system is computed non-subtractively as $\varepsilon_{\text{NS}}=Q(y)-x$. At low resolution, without dithering, $\varepsilon_{\text{NS}}$ is strongly correlated with the input signal and time-lagged versions of itself \cite{gray1993dithered}, observed by the autocorrelation vector.  

The dither $V$ follows the triangular probability density function (TPDF) defined in (\ref{eq:tpdf_a}), and is denoted as $f_{V}(v)$. In the case of no dithering, $f_V(v)=\delta(v)$, while for full dithering $f_V(v) = \Lambda_{2\Delta}(v)$. The input $X$, corresponding to a speech signal, is approximately modeled as a Laplacian probability distribution \cite{Gazor2003}, $f_X(x) = \frac{1}{2c}\exp(-\frac{|x-\mu|}{c})$ where $c$ is a scale parameter.

% \begin{equation} \label{eq:tpdf_a}
% \Lambda_{2a} \triangleq \begin{cases} 
% \frac{1}{a^2}(a + v) & -a \leq v < 0 \\ 
% \frac{1}{a^2}(a - v) & 0 < v \leq a \\ 
% 0 & \text{otherwise} 
% \end{cases}
% \end{equation}
\begin{equation} \label{eq:tpdf_a}
\Lambda_{2a}(v) \triangleq \begin{cases} \frac{1}{a^2}(a - |v|) & |v| \leq a \\ 0 & \text{otherwise} \end{cases}
\end{equation}

Though not seen with non-subtractive dithering (NSD), $\varepsilon_\text{S}$, quantization error under subtractive dithering, achieves full statistical independence from the input signal \cite{Lipshitz1992Quantization}. Despite this advantage, we neglect the method, as its implementation requires retaining the dither following ADC and results in a high-resolution output \cite{Kasher2024}; both of which contradict our low-complexity approach. However, under NSD, TPDF dithers ensure that the first and second moments of $\varepsilon_\text{NS}$ are independent from the inputs, where the second moment is minimized  \cite{Wannamaker1992NonSubtractive}. 

The total error's second moment, also referred to as the Mean Squared Error (MSE), calculated  $\mathbb{E}[\varepsilon_{\text{\text{NS}}}^2]$, is commonly used as a distortion metric, as it directly represents the power of the quantization error. With no dithering, $\mathbb{E}[\varepsilon_{\text{NS}}^2] \geq \frac{\Delta^2}{12}$, while with full dithering, $\mathbb{E}[\varepsilon_{\text{NS}}^2]\ge\frac{\Delta^2}{4}$.

\subsection*{ASR Model}

OpenAI’s \verb|Whisper| model’s front-end consists of an audio processing block that resamples all input to 16 kHz and computes a mel-spectrogram \cite{radford2022robust}, highlighting the importance of time–frequency analysis for our analysis of speech. The model's Transformer-based, end-to-end architecture is representative of the dominant frameworks in modern speech recognition \cite{bozic2024survey}; its high out-of-the-box accuracy offers a strong baseline for evaluating the effects of our parameters on a state-of-the-art system.

\Section{Analysis}
\vspace{-2mm}

\subsection*{Motivation}

Given the reliance of ASR models on feature extraction from log-mel spectrograms and prior work investigating the impact of spectral distortion on ASR performance \cite{radford2022robust}, we consider the structural integrity of Time-Frequency (TF) features ``visible'' to models as a baseline for optimal dither design. For an error signal $\varepsilon$ of total length $N$ with discrete time and frequency indices $n$ and $f$, we adopt two metrics to characterize this integrity: one quantifies the overall increase in quantization error power (MSE), and the other captures the spectral distribution of the error, both of which are represented by the Power Spectral Density (PSD) as $S_{\varepsilon\varepsilon}(f) = \left|\sum_{n=0}^N\varepsilon(n)e^{-j2\pi fn/N}\right|^2$.
% \begin{equation} \label{eq:psd}
% S_{\varepsilon\varepsilon}(f) = \lim_{T\to\infty}\frac{1}{T}\left|\sum_{n=0}^N\varepsilon_{\text{NS}}e^{-j2\pi fn/N}\right|^2 
% \triangleq \text{PSD}
% \end{equation}
% \begin{equation} \label{eq:psd}
%  % \text{PSD} \triangleq
% S_{\varepsilon\varepsilon}(f) = \left|\sum_{n=0}^N\varepsilon(n)e^{-j2\pi fn/N}\right|^2 
% \end{equation}
%S_{\varepsilon\varepsilon}(f)$ with 
%discrete frequency index $f$. %, as defined in (\ref{eq:psd}). 
We aim for a smooth error power spectrum and minimal frequency energy increase, to best ensure retention of the critical TF patterns that ASR models are trained on. 

Taking this into account, we define two independent and identically distributed (iid) dither distributions $(v \sim f_{V_{m,\alpha}} (\text{also denoted} \; f_{V_m}))$ where $V_{m,\alpha}$ is the associated random variable with choice parameter $m$. The distributions are given by (\ref{eq:ditherdef}). Controlled by the $\alpha \in [0, 1]$ parameter, $f_{V_{1,\alpha}}$ varies only in the proportion of samples allocated to the dither, while $f_{V_{2,\alpha}}$ varies both in the prior sense and with its support interval.
\vspace{-2mm}
% \begin{equation} \label{eq:ditherdef}
% \begin{split}
% f_{V_{1, \alpha}}(v)  = \alpha\Lambda_{2\Delta}(v) + (1 - \alpha) \delta(v)\\
% f_{V_{2,\alpha}}(v) = \alpha\Lambda_{2\alpha\Delta}(v) + (1 - \alpha) \delta(v)
% \end{split}
% \end{equation}
\begin{equation} \label{eq:ditherdef}
\begin{split}
f_{V_{m, \alpha}}(v)  = \alpha\Lambda_{2\Delta ((\alpha - 1)m + 2-\alpha)}(v) + (1 - \alpha) \delta(v) \quad  m \in\{1, 2\}
\end{split}
\end{equation}

\begin{figure}[h]
\centering
    \includegraphics[width=0.85\textwidth]{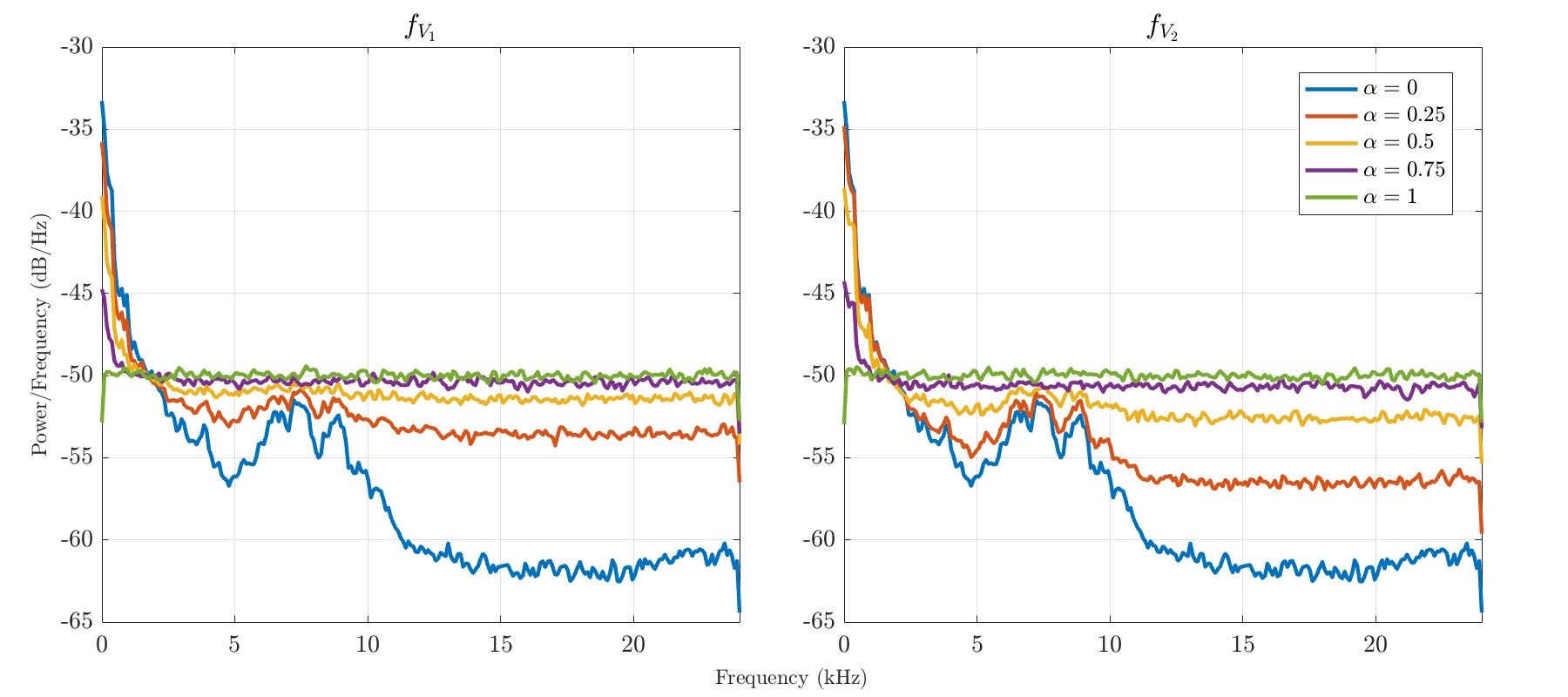} 
    \caption{PSD of $\varepsilon_{m,\alpha}$ at 1 Bit for $f_{V_{1}}$ and $f_{V_{2}}$, Given $\alpha$ = 0, 0.25, 0.5, 0.75, and 1, Smoothed With a 480-Sample Window.}
    \label{fig:psdeps}
\end{figure}

There is no difference in the impact parametric dithering has on the PSD of the quantization error signal, $\varepsilon_{m,\alpha}$, across tested bit-depths. As shown in Fig.~\ref{fig:psdeps}, null dithering concentrates power at 1 kHz, a band critical to speech \cite{Ikuma2023}, leading to dominant perceptual distortion. Increasing $\alpha$ progressively shapes the PSD of $\varepsilon_{m,\alpha}$, not only raising the noise-floor power, but also reducing the power around 1 kHz. Across $\alpha$, the PSD shape evolves non-linearly and uniquely for each dither type, with full dithering yielding a flat error PSD.

% We estimate the PSD of signal $x$ by its autocorrelation vector $r_{xx}$ in (\ref{eq:acfpsd}), where $\tau$ is the discrete time lag, referencing the Wiener-Khinchin Theorem \cite{kschischang2017wiener}; 
% \begin{equation} \label{eq:acfpsd}
% S_{xx}(\omega) = \mathcal{F}^{-1}\{r_{xx}(\tau)\}, 
% \end{equation}
For error signal $\varepsilon$, we consider specific sample index $n$ and time-lag $\tau$ with which we define autocorrelation vector $r_{\varepsilon\varepsilon}$ where $r_{\varepsilon\varepsilon}(n, \tau) = \mathbb{E}[\varepsilon_n\varepsilon_{n + \tau}] \; \text{s.t.} \; n, 
n+\tau \leq N$. Although a speech signal is not strictly WSS, on short time scales, it behaves quasi‐WSS \cite{ma2010doa} so $\forall n, r_{\varepsilon\varepsilon}(\tau) \triangleq r_{\varepsilon\varepsilon}(n, \tau)$. To measure the fine-scale temporal correlations of the quantization error, we propose $\text{ACF}_\tau$ in (\ref{eq:acf5}), where $r_{\varepsilon\varepsilon}(\tau)$ is normalized by $\mathbb{E}[\varepsilon_n^2]$ to remove the scale dependence of the absolute energy of $\varepsilon$. Our choice of $\tau =5$ reduces sensitivity to purely sample-to-sample artifacts present at $\tau=1$, yet remains short enough to capture residual correlations, including contributions from both low-frequency energy and high-frequency structures.
% whose envelopes persist across a few samples.

% \begin{equation}\label{eq:acf5}
% \text{ACF}_{\tau} \triangleq r_{\varepsilon\varepsilon}(\tau)/r_{\varepsilon\varepsilon}(0)=\mathbb{E}[\varepsilon_n\varepsilon_{n+\tau}]/\mathbb{E}[\varepsilon^2]
% \end{equation}

\begin{equation}\label{eq:acf5}
\text{ACF}_{\tau} \triangleq r_{\varepsilon\varepsilon}(\tau)/r_{\varepsilon\varepsilon}(0)=\mathbb{E}[\varepsilon_n\varepsilon_{n+\tau}]/\mathbb{E}[\varepsilon_n^2]
\end{equation}

\begin{figure}[h] 
\centering
\includegraphics[width=1\textwidth]{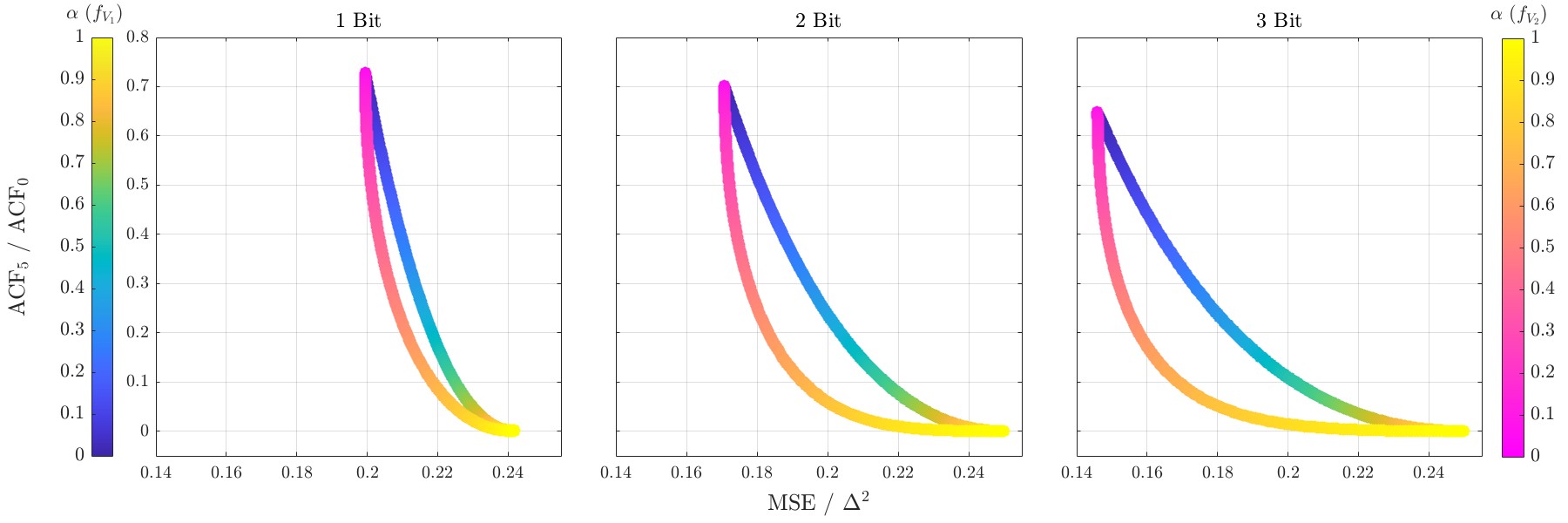}
\caption{Multi-Objective Performance Comparison between $f_{V_1}$ and $f_{V_2}$ for Tested Bit-Depths}
\label{fig:multi}
%\label{fig:optimalalpha}
\end{figure}

Increasing $\alpha$ decreases the autocorrelation of the quantization error 
%while simultaneously increasing the error's total power 
and increases the total error power, revealing a trade-off between these two metrics and motivating the use of parametric dithering. As shown in Fig. \ref{fig:multi}, 
% this trade-off between MSE and $\text{ACF}_5$ motivates the use of a parametric dither to optimize the performance of ASR. Notably, 
the Pareto front of MSE versus $\text{ACF}_5$ is different between $f_{V_1}$ and $f_{V_2}$; however, the trade-off shape is preserved across bit depths, with the curves undergoing a scaling transformation. 

Given this trade-off, we propose that (\ref{eq:propmodel}) holds as an accurate model for ASR performance, where $M(V_{m,\alpha})$, a weighted combination of MSE and $\text{ACF}_\tau$, correlates positively with $P(V_{m,\alpha})$, a direct ASR performance measure based on transcription accuracy, whose observed values vary with $\alpha$ (elaborated further in the Results section). We use this definition to numerically evaluate the trade-off coefficient, $\beta^{*}$, using (\ref{eq:relevantbeta}) over a discretized set of $\alpha$ values, where $\mu_{P} = \mathbb{E}\left[P(V_{m,\alpha})\right]$, $\sigma_{P}^{2} = \mathbb{E}\left[\left(P(V_{m,\alpha}) - \mathbb{E}\left[P(V_{m,\alpha})\right]\right)^{2}\right]$; using equivalent definitions for $M(V_{m,\alpha}, \beta)$. 

% \begin{equation} \label{eq:propmodel}
% \begin{aligned}
%    &\exists\; \beta \in [0,1]
%    \;\text{s.t.}\; \forall \alpha \in [0,1]\;
%    M(f_V) = (\beta-1)E(f_V) + \beta A(f_V) \propto P(f_V), \\
%    &E(f_V) \triangleq \mathbb{E}[\varepsilon^2], \\
%    &A(f_V) \triangleq \mathbb{E}[\varepsilon_t \varepsilon_{t+5}].
% \end{aligned}
% \end{equation}
\begin{equation} \label{eq:propmodel}
\begin{aligned}
   &\exists\; \beta \in [0,1]
   \;\text{s.t.}\; \forall \alpha
   % \in [0,1]
   \;
   M(V_{m,\alpha}, \beta) = (\beta-1)\mathbb{E}[\varepsilon_{m, \alpha}^2] + \beta \mathbb{E}[\varepsilon_{m,\alpha_{t}} \varepsilon_{m, \alpha_{t+\tau}}] \sim P(V_{m,\alpha})
\end{aligned}
\end{equation}

% This framework reflects a given direct ASR performance metric, denoted $P(V_{m,\alpha})$, within a simple dithered quantization pipeline.
%We define vectors, $\textbf{P}$ and $\textbf{M}$ where $\textbf{P}_i = P(V_{\alpha_{i}})$ and $\textbf{M}_i = M(V_{\alpha_{i}})$, to determine the appropriate scale constant, $\beta^*$ using (\ref{eq:relevantbeta}).

% \begin{equation} \label{eq:relevantbeta}
%    \beta^* = \underset{\beta}{\arg\max} \, \frac{\sum(\textbf{P}_i-\overline{\textbf{P}})(\textbf{M}_i-\overline{\textbf{M}})}{\sqrt{\sum(\textbf{P}_i-\overline{\textbf{P}})^2\sum(\textbf{M}_i-\overline{\textbf{M}})^2}} 
% \end{equation}
\begin{equation} \label{eq:relevantbeta}
   \beta^* = \underset{\beta}{\arg\max} \, \frac{\mathbb{E}\left[\left(P(V_{m,\alpha}) - \mu_{P}\right) \left(M(V_{m,\alpha},\beta) - \mu_{M}\right) \right]}{\sigma_{P}\sigma_{M}}
\end{equation}

\subsection*{Rate Computation}

In designing an efficient codec, we treat the data rate as a key factor in minimizing the required transmission bandwidth. By (\ref{eq:ditherdef}), $\alpha$ increases entropy, for both $f_{V_1}$ and $f_{V_2}$. To evaluate the difference between the entropy behaviors of $f_{V_1}$ and $f_{V_2}$, we consider Shannon's entropy, which represents the theoretical lower bound on the average number of bits per symbol required for lossless compression \cite{kulkarni2014ele201}. Shannon's entropy of output $Q_{m,\alpha}$ can be computed using (\ref{eq:quantizedentropy}), where $f_{Y_{m,\alpha}}(y) = [f_X(x)*f_{V_{m,\alpha}}(v)](y)$ and  $p_k(\alpha) = \int_{T_k}^{T_{k+1}} f_{Y_{m,\alpha}}(y)dy $.

% We start with the PDF of the input, which is the sum of two random variables $X + V = Y$, is $f_Y(y;\alpha) = [f_X(x)*f_V(v;\alpha)](y)$, where $*$ denotes the convolution operation. Following quantization, the probability of any $C_k$ is expressed as: $p_k(\alpha) = \int_{T_k}^{T_{k+1}} f_Y(y;\alpha)dy $. This can be plugged into Equation~\ref{eq:quantizedentropy} to compute entropy directly.

\begin{equation}\label{eq:quantizedentropy}
H(Q_{m,\alpha}) = -\sum_{k=1}^{2^b} p_k(\alpha) \log_2 p_k(\alpha)
\end{equation}

While a closed-form expression of Entropy could provide an exact characterization via (\ref{eq:quantizedentropy}), we gain an intuitive understanding of the dithers' behaviors from our analytical and numerical results.  

% \begin{equation} \label{eq:tpdf_var}
% \sigma_{\Lambda}^2 = \frac{a^2}{4}
% \end{equation}

We begin our analysis by considering the variance of the TPDF dithers as $\sigma_{\Lambda}^2 \triangleq a^{2}/4$, from which the variance of $V_{m,\alpha}$ follows: $\sigma_m^{2} = \text{Var} (V_{m,\alpha}) \triangleq \frac{\alpha^{2m-1}\Delta^2}{4}$.
%(\ref{eq:tpdf_var})
%the variances of $f_{V_{1}}$ and $f_{V_{2}}$ are: 
 % tpdf 1: a = $\Delta$
% tpdf 2: a = $\alpha\cdot \Delta$
% Var1 =  $\alpha \cdot \frac{(\Delta)^2}{4}$
% Var2 =  $\alpha \cdot \frac{(\alpha\cdot \Delta)^2}{4}$
% Var1 =  $\alpha \cdot \frac{(\Delta)^2}{4}$
% Var2 =  $\alpha \cdot \frac{(\alpha\cdot \Delta)^2}{4}$  
The variance of a Laplace-distributed speech input is $\text{Var}(X) \triangleq 2c^2$.
By principle of our dithers being independent from our input, $\text{Cov}(X, V_{m, \alpha}) = 0$, and therefore, $\text{Var}(X+V_{m, \alpha}) = \text{Var}(X) + \text{Var}(V_{m, \alpha})$. By Sheppard’s Corrections analysis, $\text{Var}(Q_{m,\alpha}) \approx \text{Var}(X+V_{m, \alpha}) - \frac{\Delta^2}{12}$ \cite{vardeman2005sheppard}. With this, we apply the Principle of Maximum Entropy, where the Gaussian distribution maximizes entropy for a given variance constraint \cite{conrad_n_d}. For a Gaussian variable $Z$ with variance $\sigma_Z^2$, the differential entropy is $H(Z) = \tfrac{1}{2}\log_2\!\big(2\pi e\sigma_Z^2\big)$.
While (\ref{eq:entropyresults}), which assumes a Gaussian output distribution, does not predict the exact numerical values, it captures the regression for $\alpha \in [0,1]$ and shows that $H(Q_{1,\alpha}) > H(Q_{2,\alpha})$ in that interval. 

% \begin{equation} \label{eq:entropyresults}
% \begin{split}
% H(Q_1;\alpha) &\leq \tfrac{1}{2}\log_2\!\left(2\pi e \left(2c^2 + \alpha\sigma_{1,d}^2 - \tfrac{\Delta^2}{12}\right)\right),\\
% H(Q_2;\alpha) &\leq \tfrac{1}{2}\log_2\!\left(2\pi e \left(2c^2 + \alpha^3\sigma_{2,d}^2 - \tfrac{\Delta^2}{12}\right)\right).
% \end{split}
% \end{equation}
\begin{equation} \label{eq:entropyresults}
H(Q_{m,\alpha}) \leq \tfrac{1}{2}\log_2\!\left(2\pi e \left(2c^2 + \sigma_m^2 - \tfrac{\Delta^2}{12}\right)\right)
\end{equation}

\begin{figure}[!h] %\label{fig:entropy}
\centering
\includegraphics[width=1.05\textwidth]{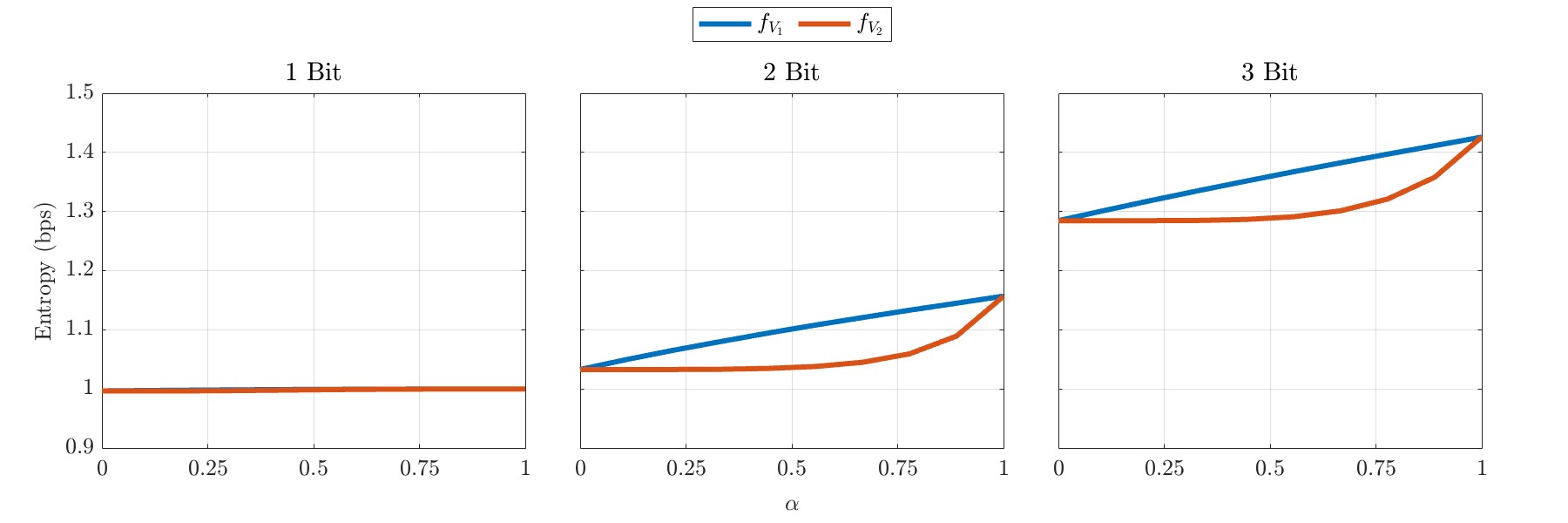}\\
\caption{Numerical Results for Shannon's Entropy Over $\alpha$ for Speech Quantized to 1-3 Bits}
\label{fig:entropy}
\end{figure}

Fig.~\ref{fig:entropy} numerically validates (\ref{eq:entropyresults}): entropy increases with $\alpha$, while the trend differs significantly between the two dither types. Entropy for $f_{V_{1}}$ exhibits a concave shape over alpha, while with $f_{V_{2}}$, entropy shows a distinctly convex trend. This convex behavior, a result of the controlled dither support, demonstrates the advantage of $f_{V_2}$ dithering.

\Section{Codec}
\vspace{-2mm}

\begin{figure}[h] 
\centering
\includegraphics[width=0.85\textwidth]{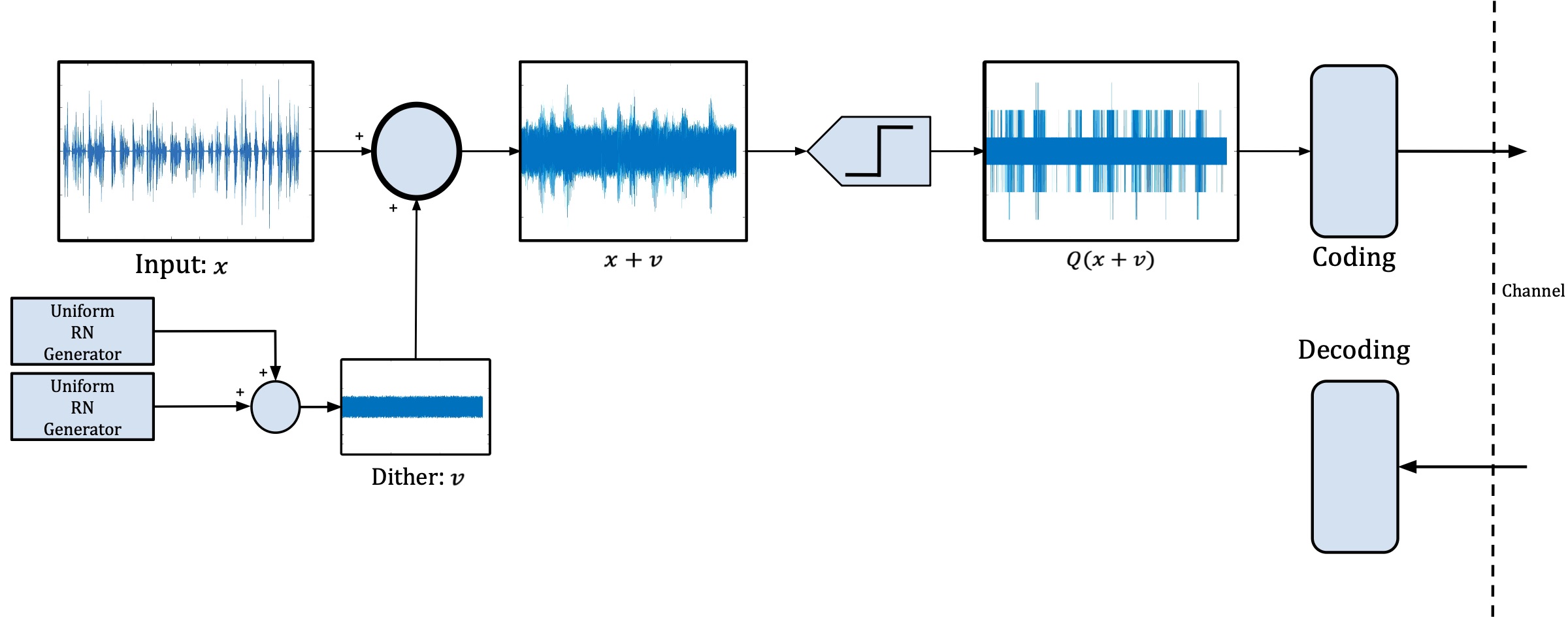} 
    \caption{Speech Codec Quantization Pipeline}
\label{fig:pipeline}\vspace{-4mm}
\end{figure}

The proposed codec follows the pipeline outlined in Fig. \ref{fig:pipeline}. 
% Dither generation is implemented by summing the outputs of Uniform Random Number Generators. 
Input and dither signals of $N$ samples are summed before being passed through the mid-tread $b$-bit quantizer. 
This is followed by 
Huffman Coding, an optimal lossless compression technique
%, follows, achieving 
which achieves
an average code length 
%$L$, where $H(X) \le L \le H(X) +1$, with $H(X)$ being Shannon's entropy of the source.
within 1 bit of the source entropy.

%\vspace{-2mm}

\Section{Results}
\vspace{-2mm}

\subsection*{Experimental Setup}
We used 30 identical scripts, distinct voices' speech samples drawn from Google AI’s \verb|en-US-Chirp3-HD| voices \cite{googlechirp3}. All files (16-bit linear PCM, 48 kHz) were imported into MATLAB, where they were amplitude-normalized and trimmed to 20 seconds. Our intelligibility metrics were calculated on the decoded signals, and our rate calculation uses Huffman coding. 

We utilize \verb|whisper.cpp|, a high-performance C++ implementation of the original model \cite{whispercpp}. Whisper’s transcription accuracy is evaluated by the Character Error Rate (CER), computed via the Levenshtein distance. The reference transcript is taken from the high-quality imported speech itself, rather than a global key, to account for variable content across the clipped samples. The Levenshtein distance is normalized by the reference text length, providing a theoretical score in the range [0, 1], corresponding to  $P(V_{m, \alpha}) \triangleq \text{CER}$. To produce a single data point given each bit-resolution and $\alpha$ parameter while accounting for inter-speaker variability, for each metric, we computed the mean and the standard error of the mean (SEM).

\subsection*{ASR Performance}
\begin{figure}[h] 
\centering
\includegraphics[width=1.04\textwidth]{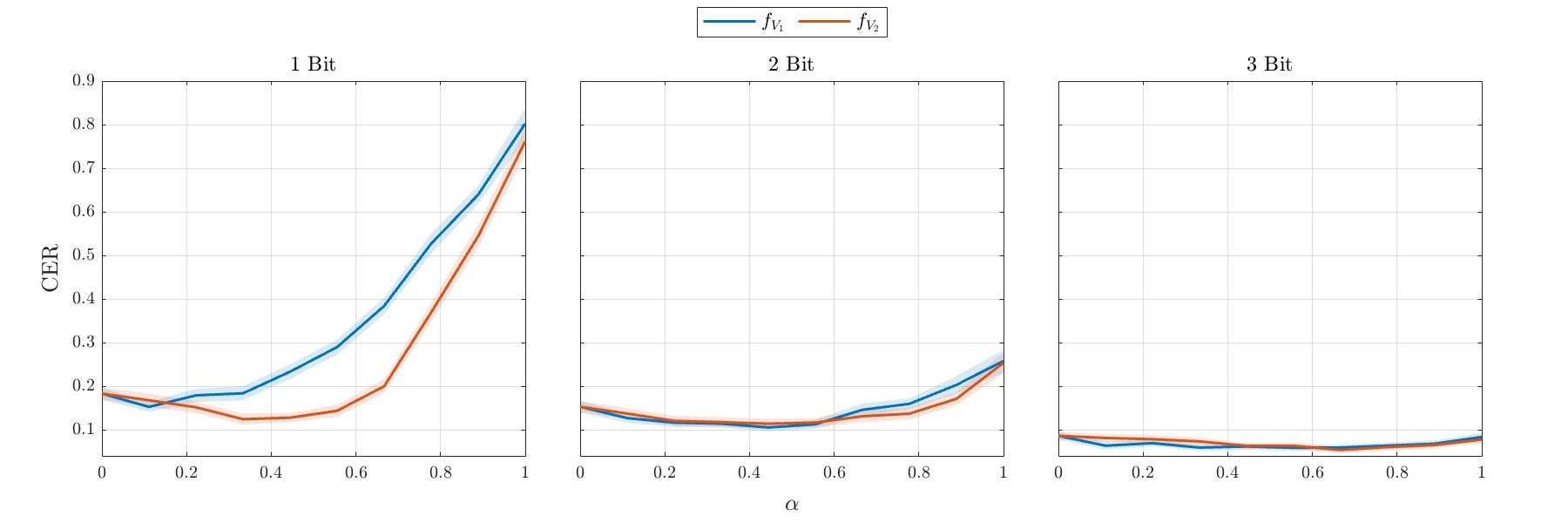}
\caption{$P(V_{1, \alpha})$ and $P(V_{2, \alpha})$ Results Across $\alpha$; Shaded Regions Indicate the SEM}\vspace{-4mm}
%\label{fig:metrics}
\label{fig:cervalpha}
\end{figure}

As shown in Fig.~\ref{fig:cervalpha}, $P(V_{m, \alpha})$ is minimized by parametric dithering for all tests. 
Despite the lower overall $P(V_{m, \alpha})$ at 3-bit resolution (maximum absolute improvement = 0.0271), the largest absolute gains occur at 1- and 2-bit quantization (0.0449 and 0.0486 improvement), suggesting weaker motivation at the higher resolution case. 

\vspace*{-2mm}

\subsection*{Model Verification}
\begin{figure}[h] 
\centering
\includegraphics[width=1.05\textwidth]{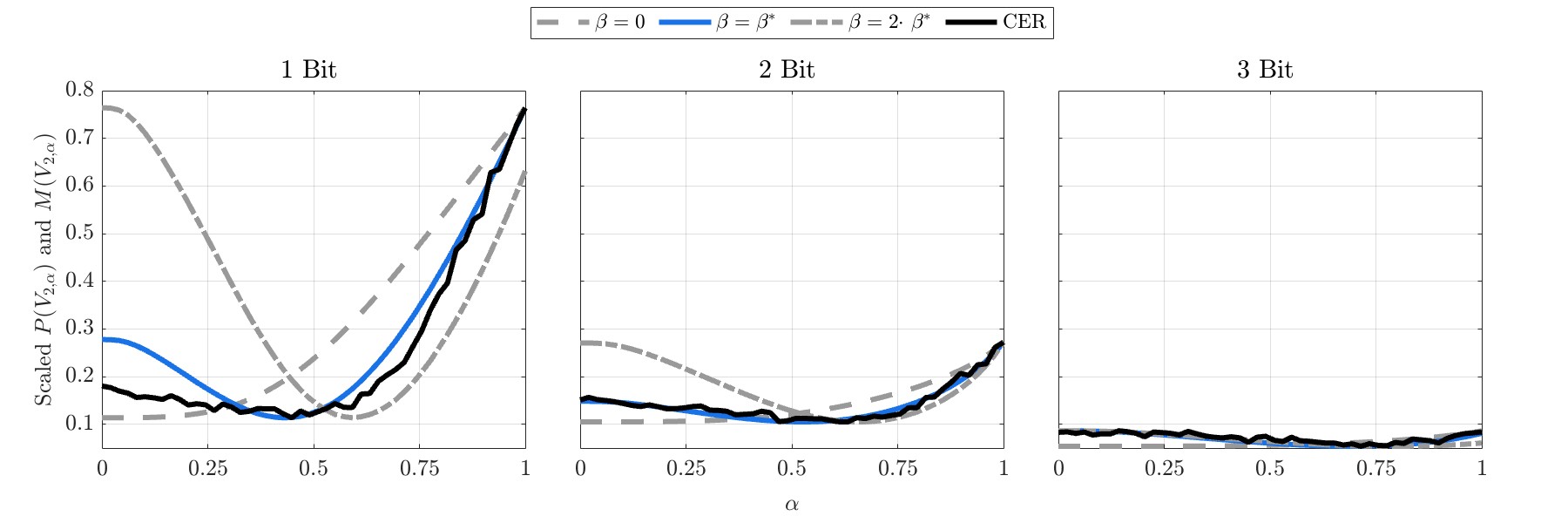}
\caption{Scaled $M(V_{2,\alpha})$ Within the Range of $P({V_{2,\alpha}})$ and Fitted to $P(V_{2,\alpha})$ for $\beta^*$ and Local $\beta$ Values.}

\label{fig:corrplot}
\end{figure}

The verification of metric $M(V_{m, \alpha})$ is verified using the method prescribed by (\ref{eq:propmodel}). The $\beta^*$ that satisfy the model are shown in Table \ref{tab:relevantbeta}, where $\beta^*$ is shown to vary across bit-depth, and less significantly between the two dither types. Total power of $\varepsilon_{\text{NS}}$ (MSE) holds a substantially larger weight in $M(V_{m, \alpha})$ than the shape of noise floor ($\text{ACF}_5$); both qualities are relevant. Notably, $\text{ACF}_5$ holds more weight at higher rates. Fig. \ref{fig:corrplot} visualizes the fit of $M(V_{m, \alpha})$ to the results of $P(V_{m, \alpha})$.

\begin{table}[h!] 
    \centering 
    \caption{Computed $\beta^*$ Values}
    \label{tab:relevantbeta}
    \label{tab:dither_distributions}
    \begin{tabular}{cccc}
        \toprule
        Bit-Depth & 1 & 2 & 3\\
        \midrule
        $f_{V_1}$ &0.0303 & 0.0808 & 0.141\\
        $f_{V_2}$ &0.0303 & 0.0505 & 0.152\\
        \bottomrule
    \end{tabular}
\end{table}

\subsection*{Optimal Alpha}

\begin{figure}[h] 
\centering
\includegraphics[width=0.55\textwidth]{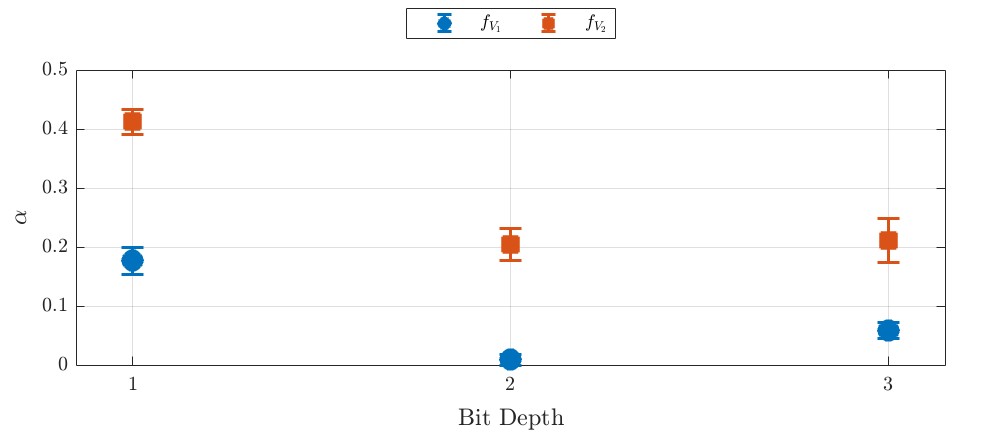}
\caption{Computed $\alpha^*$ Values with SEM Error Bars}\vspace{-4mm}
\label{fig:optimalalpha}
\end{figure}

The optimal $\alpha$ value, $\alpha^*$, is determined by evaluating both CER improvement and rate. However, for 1-bit quantization, the entropy remains constant, making CER the sole determining factor for optimization. The $\alpha^*$ computed by (\ref{eq:optimalalpha}), where $R(\alpha)$ denotes the rate (Huffman coding) given $\alpha$ and $P(\alpha) = P(V_{m,\alpha}(\alpha))$, are shown in Fig.~\ref{fig:optimalalpha}. The $\alpha^*$ for $f_{V_2}$ is consistently larger than that of $f_{V_1}$ even at similar CER improvement, reflecting the reduced rate inherent to $f_{V_2}$.

% \begin{equation} 
% \label{eq:optimalalpha}
% \begin{split}
%    \alpha^{*} = \underset{\alpha}{\arg\max} \, \frac{C(0) -C(\alpha)}{H(0) -H(\alpha)} \quad \text{for} \quad \{ \alpha \mid C(\alpha) < C(0) \}
% \end{split}
% \end{equation}
\begin{equation} 
\label{eq:optimalalpha}
\begin{split}
   \alpha^{*} = \underset{\alpha :\, P(\alpha) < P(0)}{\arg\max} \, \frac{P(0) -P(\alpha)}{R(0) -R(\alpha)}% \quad \text{for} \quad \{ \alpha \mid C(\alpha) < C(0) \}
\end{split}
\end{equation}

\subsection*{Rate Distortion}

\begin{figure}[h] %\label{fig:ratedistortion}
\centering
\includegraphics[width=1.1\textwidth]{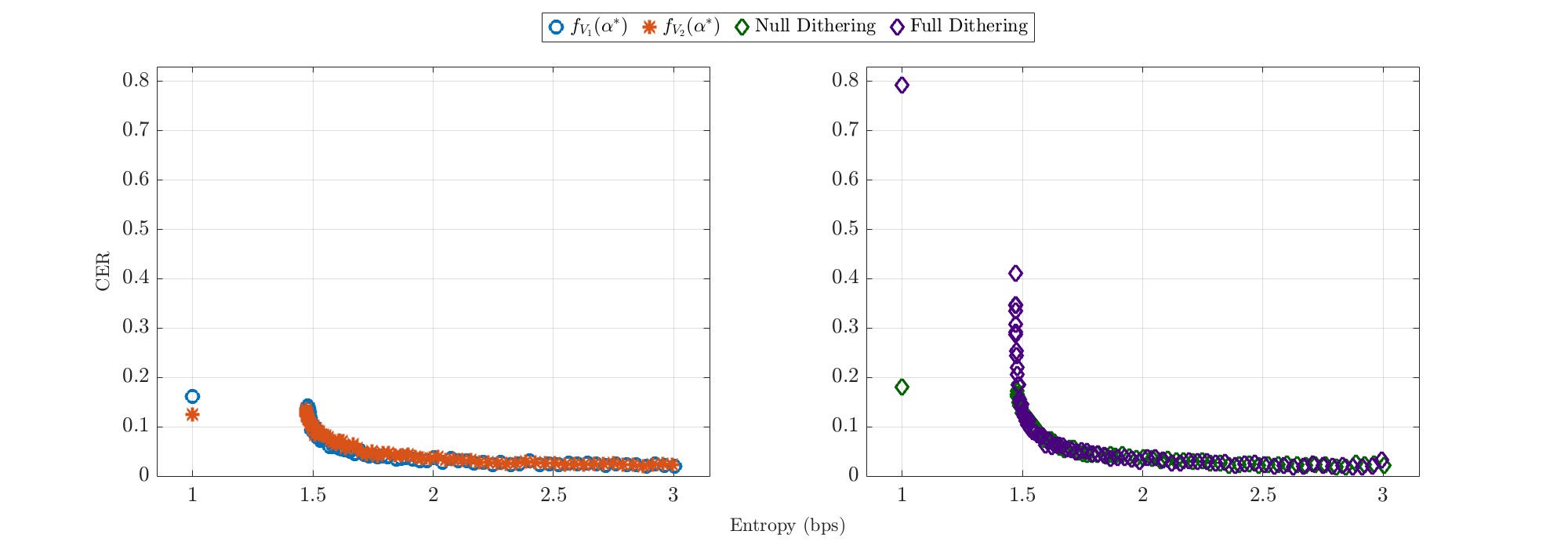}
\caption{(a) CER of $f_{V_1}(\alpha^*)$ and $f_{V_2}(\alpha^*)$ Dithering Compared to (b) CER of Null and Full Dithering as a Function of Entropy}
\label{fig:ratedistortion}
\end{figure}

Fig. \ref{fig:ratedistortion} shows that at 1-bit quantization, $f_{V_2}(\alpha^*)$ dithering yields a lower CER than $f_{V_1}(\alpha^*)$, showing clear dominance under the most extreme entropy constraint. While increasing the quantization resolution improves ASR performance, it also raises the data rate and provides diminishing returns in CER improvement. As the rate continues to increase, $f_{V_1}(\alpha^*)$ dithering eventually matches and slightly surpasses $f_{V_2}(\alpha^*)$, until their performances become indistinguishable. Notably, at 1-bit quantization, $f_{V_2}(\alpha^*)$ is most justified: not only outperforming null and full dithering by a significant margin, but also achieving minimal rate.

% \Section{Discussion}

% These results are promising in the design of low-power, low-latency hardware implementations aimed at real-time speech compression, while maintaining quality for ASR input. The approach relies on simple digital dither generation, which can be realized by summing two LUT-based pseudo-random number generators, such as FPGA-optimized LUT-Shift Registers \cite{ThomasLuk_LUTSR} paired with an R-2R Digital-to-Analog Converter, and quantization implemented by a flash ADC \cite{balasubramanian1995flash}. In such systems, processing speed and latency are determined largely by the analog front-end, enabling effective use in systems that rely on offloading for complex functions \cite{ha2013towards}. 

% Standard codecs such as Opus provide highly efficient, near-lossless compression, even at low rates, but rely on FFT-based computations \cite{vos2013opus}. These computations demand significant CPU resources or specialized DSP hardware, which can reduce system responsiveness and increase latency. Platforms such as wearable devices can benefit from the efficiency of our approach, which in turn enables these devices to reduce communication barriers for users with hearing impairments or in multilingual environments.

\vspace{-2mm}
\Section{Conclusion}
\vspace{-2mm}

% We proposed the design of a low-complexity, entropy-constrained speech compression codec optimized for ASR. 
We propose a low-complexity, entropy-constrained speech compression codec design optimized for ASR. We introduced a framework for characterizing ASR performance using MSE and $\text{ACF}_5$, providing a quantitative basis for designing and evaluating dithering strategies. Our experiments demonstrate the effectiveness of 1-bit quantization with parametric dithering, achieving a critically low CER and yielding a 25\% relative improvement compared to null-dithering. These results highlight the potential of ultra–low bit-depth quantization as a viable approach for speech waveform coding in resource-constrained systems.

Future directions include retraining a transformer-based encoder–decoder ASR model to evaluate the limits of parametric dithering more fully. Additional work may focus on expanding the ASR performance model through broader testing and consideration of supplementary objective measures, further validating the role of dithering in ultra-low-resolution speech coding.

\Section{References}
\vspace{-2mm}
\bibliographystyle{IEEEbib}
\bibliography{refs}

@article{gray1998quantization,
  author  = {Gray, R. M. and Neuhoff, D. L.},
  title   = {Quantization},
  journal = {IEEE Transactions on Information Theory},
  year    = {1998},
  volume  = {44},
  number  = {6},
  pages   = {2325--2383},
  month   = {Oct.},
}

@misc{vanderkooy1987dither,
  author = {Vanderkooy, J. and Lipshitz, S. P.},
  title  = {Dither in digital audio},
  year   = {1987},
  note   = {[Online]. Available: \url{https://aes2.org/publications/elibrary-page/?id=5173}},
}

@techreport{brandenburgMP3,
  author      = {Brandenburg, K.},
  title       = {{MP3} AND {AAC} {EXPLAINED}},
  institution = {Fraunhofer Institute for Integrated Circuits},
  year        = {n.d.},
  address     = {Erlangen, Germany},
}

@article{lo2021architectural,
  author = {S. K. Lo and Q. Lu and L. Zhu and H.-y. Paik and X. Xu and C. Wang},
  title = {Architectural Patterns for the Design of Federated Learning Systems},
  journal = {Journal of Systems and Software},
  note = {Special issue on Software Architecture and Artificial Intelligence},
  year = {2021},
  howpublished = {Submitted on 7 Jan 2021, revised 18 Jun 2021},
  publisher = {Elsevier}
}

@inproceedings{Freitas2011,
  author    = {P. Freitas and M. Farias and A. Araujo},
  title     = {Fast Inverse Halftoning Algorithm for Ordered Dithered Images},
  booktitle = {2011 24th SIBGRAPI Conference on Graphics, Patterns and Images},
  address   = {Alagoas, Brazil},
  year      = {2011},
  pages     = {250--257},
  doi       = {10.1109/SIBGRAPI.2011.14}
}

@inproceedings{borsky_spectrally,
  author       = {Michal Borsky and Petr Mizera and Petr Pollak},
  title        = {Spectrally Selective Dithering for Distorted Speech Recognition},
  booktitle    = {InterSpeech}, 
    year         = {2015}, 
    institution  = {Faculty of Electrical Engineering, Czech Technical University in Prague, Czech Republic},
  note         = {Emails: borskmic@fel.cvut.cz, mizerpet@fel.cvut.cz, pollak@fel.cvut.cz}
}

@article{borsky2015advanced,
  author = {M. Borsky and P. Pollak and P. Mizera},
  journal = {EURASIP Journal on Audio, Speech, and Music Processing},
  number = {1},
  pages = {20},
  title = {Advanced acoustic modelling techniques in MP3 speech recognition},
  volume = {2015},
  year = {2015},
  doi = {10.1186/s13636-015-0052-0},
  url = {https://doi.org/10.1186/s13636-015-0052-0}
}

@article{borsky2017dithering,
  author = {Michal Borsky and Petr Mizera and Petr Pollak and Jan Nouza},
  title = {Dithering techniques in automatic recognition of speech corrupted by MP3 compression: Analysis, solutions and experiments},
  journal = {Speech Communication},
  volume = {86},
  pages = {75--84},
  year = {2017},
  issn = {0167-6393}
}

@article{gray1993dithered,
  author  = {Gray, R. M. and Stockham, T. G.},
  title   = {Dithered quantizers},
  journal = {IEEE Transactions on Information Theory},
  year    = {1993},
  volume  = {39},
  number  = {3},
  pages   = {805--812},
  month   = {May},
  doi     = {10.1109/18.256489},
}

@article{Lipshitz1992Quantization,
  author  = {Stanley P. Lipshitz and Robert A. Wannamaker and John Vanderkooy},
  title   = {Quantization and Dither: A Theoretical Survey},
  journal = {Journal of the Audio Engineering Society},
  year    = {1992},
  volume  = {40},
  number  = {5},
  pages   = {355--375}
}

@article{Kasher2024,
  author    = {Morriel Kasher and Michael Tinston and Predrag Spasojevic},
  title     = {{Distortion-Controlled Dithering with Reduced Recompression Rate}},
  journal   = {arXiv e-prints},
  year      = {2024},
  month     = {feb},
  eprint    = {2402.17029},
  archivePrefix = {arXiv},
  primaryClass = {eess.SP}
}

@article{Wannamaker1992NonSubtractive,
  author  = {Robert A. Wannamaker and Stanley P. Lipshitz and John Vanderkooy and J. Nelson Wright},
  title   = {A Theory of Non-Subtractive Dither},
  journal = {Journal of the Audio Engineering Society},
  year    = {1992},
  volume  = {40},
  number  = {7/8},
  pages   = {571--583}
}

@article{Gazor2003,
  author    = {S. Gazor and Wei Zhang},
  title     = {Speech probability distribution},
  journal   = {IEEE Signal Processing Letters},
  volume    = {10},
  number    = {7},
  pages     = {204--207},
  month     = jul,
  year      = {2003},
  doi       = {10.1109/LSP.2003.813679}
}

@article{radford2022robust,
  title={Robust Speech Recognition via Large-Scale Weak Supervision},
  author={Radford, Alec and Kim, Jong Wook and Xu, Tao and Brockman, Greg and McLeavey, Christine and Sutskever, Ilya},
  journal={arXiv preprint arXiv:2212.04356},
  year={2022},
  url={https://arxiv.org/abs/2212.04356}
}

@misc{bozic2024survey,
  author       = {Božić, M. and Horvat, M.},
  title        = {A survey of deep learning audio generation methods},
  year         = {2024},
  howpublished = {arXiv preprint},
  note         = {arXiv:2405.20146. [Online].}
}

@article{Ikuma2023,
  author  = {T. Ikuma and A. J. McWhorter and E. Oral and M. Kunduk},
  title   = {Formant-Aware Spectral Analysis of Sustained Vowels of Pathological Breathy Voice},
  journal = {Journal of Voice},
  year    = {2023},
  issn    = {0892-1997},
  doi     = {10.1016/j.jvoice.2023.05.002}
}

@article{ma2010doa,
  author={Ma, Wing-Kin and Hsieh, Tsung-Han and Chi, Chong-Yung},
  journal={IEEE Transactions on Signal Processing},
  title={DOA estimation of quasistationary signals with less sensors than sources and unknown spatial noise covariance: A khatri--rao subspace approach},
  year={2010},
  volume={58},
  number={4},
  pages={2168-2180},
  doi={10.1109/TSP.2010.2040944},
  month={April}
}

@misc{kulkarni2014ele201,
  author = {Kulkarni, S. R.},
  title  = {{Ele 201: Information signals - course notes, chapter 8: Information, entropy, and coding}},
  year   = {2014},
}

@article{vardeman2005sheppard,
  author={Vardeman, S. B.},
  journal={IEEE Transactions on Instrumentation and Measurement},
  title={Sheppard's correction for variances and the "quantization noise model"},
  year={2005},
  volume={54},
  number={5},
  pages={2117-2119},
  doi={10.1109/TIM.2005.853348},
  month={Oct}
}

@misc{conrad_n_d,
  author = {Conrad, Keith},
  title = {PROBABILITY DISTRIBUTIONS AND MAXIMUM ENTROPY},
  howpublished = {Expository papers},
  year = {n.d.},
  note = {Accessed on 2025-08-22},
  url = {https://kconrad.math.uconn.edu/blurbs/analysis/entropypost.pdf}
}

@misc{googlechirp3,
  author       = "{GoogleCloudPlatform}",
  title        = "{get\_started\_with\_chirp\_3\_hd\_voices.ipynb -- Generative AI Audio Example}",
  year         = {2025},
  howpublished = "GitHub repository",
  note         = "[Online]. Available: \url{https://github.com/GoogleCloudPlatform/generative-ai/blob/main/audio/speech/getting-started/get_started_with_chirp_3_hd_voices.ipynb} [Accessed: Sep. 24, 2025]"
}

@misc{whispercpp,
  author       = "{ggml-org}",
  title        = "{whisper.cpp: A high-performance, dependency-free port of OpenAI's Whisper model}",
  year         = {2025},
  howpublished = "GitHub repository",
  note         = "[Online]. Available: \url{https://github.com/ggml-org/whisper.cpp} [Accessed: Sep. 24, 2025]"
}

@techreport{ha2013towards,
  author      = {Ha, K. and Chen, Z. and Hu, W. and Richter, W. and Pillai, P. and Satyanarayanan, M.},
  title       = {Towards Wearable Cognitive Assistance},
  institution = {Sch. of Comput. Sci., Carnegie Mellon Univ.},
  year        = {2013},
  number      = {CMU-CS-13-134},
  address     = {Pittsburgh, PA},
  month       = {Dec.},
}

@inproceedings{vos2013opus,
  author    = {Vos, K. and Sørensen, K. V. and Jensen, S. S. and Valin, J.-M.},
  title     = {The Opus Codec},
  booktitle = {135th AES Convention},
  year      = {2013},
  address   = {New York, USA},
}

% \Section{Appendix}

\end{document}